\documentclass[11pt,a4paper]{article}
\usepackage{jheppub}

\usepackage[retainorgcmds]{IEEEtrantools}
\usepackage{float}
\usepackage{amsmath, bm,mathtools} 
\usepackage{amssymb}
\usepackage{subcaption}
\usepackage{dsfont}
\usepackage{hyperref}
\usepackage{dsfont}
\usepackage{slashed} 
\usepackage{mathtools} 

\usepackage{tikz}
\usepackage{tkz-euclide}
\usetikzlibrary{decorations.pathmorphing}	
\tikzset{
    v/.style={decorate, decoration={snake, segment length=3mm, amplitude=0.75mm}, draw},
    f/.style={draw,decoration={markings,mark=at position #1 with {\arrow[very thick]{latex}}},postaction={decorate},node contents=#1},
    f/.default=.6,
    fb/.style={draw,decoration={markings,mark=at position #1 with {\arrowreversed[very thick]{latex}}},postaction={decorate},node contents=#1},
    fb/.default=.4,
    fnar/.style={draw},
    g/.style={decorate, draw,  decoration={coil,amplitude=3pt, segment length=3.5pt}},
    s/.style={dashed,draw, postaction={decorate},
        decoration={markings,mark=at position .55 with {\arrow[very thick]{latex}}}},
    sb/.style={dashed,draw, postaction={decorate},
        decoration={markings,mark=at position .55 with {\arrowreversed[draw=black,very thick]{latex}}}},
    snar/.style={dashed,draw,line width =1.25pt},
}
\usetikzlibrary{shapes}	
\tikzset{every picture/.style={line width=1}}

\newcounter{qnumber}

\definecolor{c1}{rgb}{0.121569, 0.466667, 0.705882}
\definecolor{c2}{rgb}{1., 0.498039, 0.054902}
\definecolor{c3}{rgb}{0.172549, 0.627451, 0.172549}
\definecolor{c4}{rgb}{0.839216, 0.152941, 0.156863}
\definecolor{c5}{rgb}{0.580392, 0.403922, 0.741176}
\definecolor{c6}{rgb}{0.54902, 0.337255, 0.294118}
\definecolor{c7}{rgb}{0.890196, 0.466667, 0.760784}
\definecolor{c8}{rgb}{0.498039, 0.498039, 0.498039}
\definecolor{c9}{rgb}{0.737255, 0.741176, 0.133333}
\definecolor{c10}{rgb}{0.0901961, 0.745098, 0.811765}

\DeclareMathOperator{\g}{\gamma}          

\usepackage{color}
\usepackage{tcolorbox}
\newtcolorbox{mybox}{colback=blue!5!white,colframe=blue!75!black}

\begin{document}

\title{QCD Axion-Mediated Dark Matter}

\author[a]{Jeff A. Dror,}
\emailAdd{jdror1@ucsc.edu} %

\author[a,b,c]{Stefania Gori,}
\emailAdd{sgori@ucsc.edu} %

\author[a]{Pankaj Munbodh}
\emailAdd{pmunbodh@ucsc.edu} %

\affiliation[a]{Department of Physics, University of California Santa Cruz, 1156 High St., Santa Cruz, CA 95064, USA\\
and Santa Cruz Institute for Particle Physics, 1156 High St., Santa Cruz, CA 95064, USA
}
\affiliation[b]{Ernest Orlando Lawrence Berkeley National Laboratory,
University of California, Berkeley, CA 94720, USA}

\affiliation[c]{Berkeley Center for Theoretical Physics, Department of Physics,
University of California, Berkeley, CA 94720, USA}

\date{\today}
\abstract{
A QCD axion with a decay constant below $ 10 ^{ 11} ~{\rm GeV}   $ is a strongly-motivated extension to the Standard Model, though its relic abundance from the misalignment mechanism or decay of cosmic defects is insufficient to explain the origin of dark matter. Nevertheless, such an axion may still play an important role in setting the dark matter density if it mediates a force between the SM and the dark sector. In this work, we explore QCD axion-mediated freeze-out and freeze-in scenarios, finding that the axion can play a critical role for setting the dark matter density. Assuming the axion solves the strong CP problem makes this framework highly predictive, and we comment on experimental targets.
}
\maketitle

\section{Introduction}
The QCD axion has long been appreciated as a solution to the strong CP problem~\cite{Peccei:1977hh,Peccei:1977ur,Weinberg:1977ma,Wilczek:1977pj} and is a common prediction of string theory~\cite{Svrcek:2006yi,Arvanitaki:2009fg,Halverson:2019cmy}. Shortly after its inception, it was realized that the QCD axion can be also a good Dark Matter (DM) candidate~\cite{Abbott:1982af,Preskill:1982cy,Dine:1982ah}. This has led to a worldwide experimental program to target axion DM with different interactions across a broad range of masses (see Ref.~\cite{Chadha-Day:2021szb} for a recent review).

While the QCD axion has built-in mechanisms to explain the observed DM abundance through the misalignment mechanism and cosmic defect decays, these mechanisms only efficiently produce axions for sufficiently large values of the decay constant, $ f _a $. Axions with $ f _a  \lesssim  10 ^{ 11} ~{\rm GeV} $ do not constitute a sizable fraction of the DM energy density without additional dynamics (see, e.g., Ref.~\cite{Co:2017mop}). Nevertheless, there is a large-scale program to detect such axions with the upcoming IAXO~\cite{IAXO:2019mpb} and ALPS II~\cite{Bahre:2013ywa} experiments. Furthermore, there have even been some tantalizing hints of detection in the star cooling data~\cite{Giannotti:2015dwa} and an excess of events reported by the XENON1T experiment~\cite{XENON:2020rca}. In this paper, we explore models that feature a low $f_a$ QCD axion mediating the interactions between DM particles and the Standard Model (SM). 

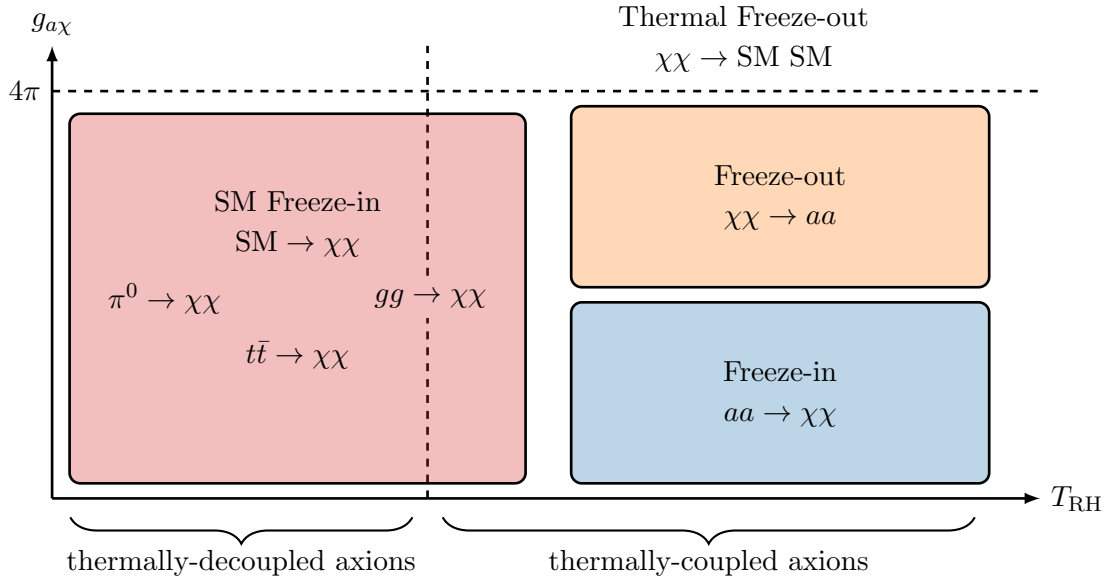
\begin{figure} 
\begin{center} \begin{tikzpicture} 
\draw[-latex] (0,0) coordinate (O) --++ (13,0) coordinate (E)node[right] {$ T _{ {\rm RH}} $};
\draw[-latex] (O) --++(0,6) coordinate (T)node[above]{$g _{a\chi} $} ;
\draw[draw=none] ($ (O)!0.51!(E)$) coordinate (split-bottom) --++(0,6) coordinate (split-top); 
\draw[dashed] ($ (O)!0.38!(E)$) coordinate (split-bottom2) --++(0,6) coordinate (split-top2);
\def\h{2.4}
\draw[rounded corners,preaction={fill=c1,opacity=0.3}] ($ (split-bottom)+(0.2,0.2) $) rectangle ++(5.5,\h) node[midway] {\begin{tabular}{c}Freeze-in \\ $ a a \rightarrow \chi \chi $ \end{tabular}};
\draw[rounded corners,preaction={fill=c2,opacity=0.3}] ($ (split-bottom)+(0.2,0.4+\h) $) rectangle ++(5.5,\h) node[midway] {\begin{tabular}{c}Freeze-out \\ $ \chi \chi  \rightarrow a a  $ \end{tabular}};
\draw[dashed] ($ (O)!0.9!(T)$) node[left] {$ 4\pi  $}coordinate (pert-left) --++(13,0) coordinate (pert-right);
\node at ($ (pert-left)!0.7!(pert-right)+(0,.7) $){\begin{tabular}{c} Thermal Freeze-out \\  $ \chi \chi  \rightarrow {\rm SM}~{\rm SM}$ \end{tabular}}; 
\draw[thick, decorate,decoration={brace,amplitude=10,mirror}] ($ (split-bottom2)+(0.2,-0.25) $) --++ (7,0)node[midway,yshift=-0.6cm] {thermally-coupled axions} ;
\draw[thick, decorate,decoration={brace,amplitude=10}] ($ (split-bottom2)+(-0.2,-0.25) $) --++ (-4.5,0)node[midway,yshift=-0.6cm] {thermally-decoupled axions};
\draw[rounded corners,preaction={fill=c4,opacity=0.3}] ($ (split-bottom)+(-0.4-0,0*\h+.2) $) rectangle ++(-6,2*\h+0.1)coordinate[midway] (C) node[midway,yshift=1cm] {\begin{tabular}{c}SM Freeze-in \\ $ {\rm SM } \rightarrow \chi \chi $\end{tabular}} ;
\node at ($ (C)+(-1.75,0) $) {$ \pi ^0 \rightarrow \chi \chi $};
\node at ($ (C)+(0,-.75) $) {$ t \bar{t}  \rightarrow \chi \chi $};
\node[preaction={fill=c4,preaction={fill=white},opacity=0.3}] at ($ (C)+(1.75,0) $) {$ gg  \rightarrow \chi \chi $};
  \end{tikzpicture}
\end{center}
\caption{Schematic outline of possible thermal histories of QCD axion-mediated dark matter as a function of the reheating temperature of the universe, $ T _{ {\rm RH}} $, and the axion-dark matter coupling, $ g _{ a \chi } $, defined below Eq. (\ref{eq:aDM}). For sufficiently high $ T _{ {\rm RH}} $, the QCD axion comes into thermal equilibrium with the SM. In this limit, $ \chi $ can freeze-out or freeze-in through the QCD axion, while conventional freeze-out to the SM is ruled out by perturbativity of $ g _{ a \chi } $. For smaller $ T _{ {\rm RH}}$, the axion does not come into thermal contact with the SM, but $ \chi $ can still be produced through direct SM freeze-in. Due to its ultraviolet nature, gluon-driven freeze-in can contribute significantly to the abundance. This can be the dominant production mechanism independently of whether the axion is thermally coupled to the SM.}
\label{fig:history}
\end{figure}

To study the viable thermal histories for QCD axion-mediated DM scenarios, we consider a Dirac fermion DM candidate, $ \chi $. The cosmology depends critically on the values of the reheating temperature of the universe after inflation ($ T _{ {\rm RH}} $). If $ T _{ {\rm RH}} $ is large enough for the axion to reach thermal equilibrium with the SM sector at early times, then $ \chi $ can also come into thermal equilibrium through axion-$ \chi $ interactions. In this case, $ \chi $ can undergo freeze-out. While QCD axion-mediated annihilations of $ \chi $ into SM particles (such as gluons or quarks) are already experimentally constrained to be too small for $ \chi $ to reach the observed abundance of DM, $\chi$ can freeze-out into axions. The axion would then remain today as dark radiation. For lower reheating temperatures, the axion and, therefore, $ \chi $ never come into equilibrium with the Standard Model. In this case,  $ \chi $ can be frozen-in either through Standard Model collisions or axion annihilations. A summary of the different thermal histories is shown in Fig.~\ref{fig:history}.

We study these cosmic histories, finding the parameter space for which $ \chi $ makes up DM and its viability in light of the relevant astrophysical and cosmological constraints. The thermal history for axion-like particle-mediated DM has been recently considered in Refs.~\cite{Bharucha:2022lty,Ghosh:2023tyz}. While some of the phenomenology we discuss in this paper overlaps with Refs.~\cite{Bharucha:2022lty, Ghosh:2023tyz}, assuming that the axion solves the strong CP problem dramatically shapes the phenomenology. We emphasize key differences throughout. In particular, the QCD axion is a highly predictive theory, and, as such, our results can be represented in terms of three model parameters: the axion decay constant ($ f _a $), the DM mass ($ m _\chi $), and the coupling of the axion to the dark sector ($ g _{a\chi} $), and one cosmological parameter namely the reheating temperature ($T_{\text{RH}}$). 

The paper is structured as follows. In Sec.~\ref{sec:model}, we introduce the Lagrangian of our model.  
In addition, we present the constraints on the Lagrangian parameters from astrophysical and cosmological data independent of the assumption that $ \chi $ makes up the full abundance for DM. In Sec.~\ref{sec:history}, we present the outline of the thermal history. In Sec.~\ref{sec:coupled}, we explore the different possible freeze-out and freeze-in thermal histories, pointing out interesting experimental targets. We summarize our results and conclude in Sec.~\ref{sec:conclusions}. 

\section{The QCD Axion and Dark Matter}
\label{sec:model}

\subsection{Introduction to the Model}
We begin by presenting the structure of the PQ symmetry, ${\rm U}(1) _{ {\rm PQ}} $, and the corresponding pseudo-Goldstone boson after spontaneous symmetry breaking, the QCD axion ($ a $). 

The fundamental interaction of the QCD axion to the gluon field strength tensor ($ G _{\mu\nu} ^A $) and its dual ($ \tilde{G}^A_{\mu\nu} = \frac{1}{2}\epsilon_{\mu\nu\rho\sigma} G^{A\rho\sigma} $) is given by
\begin{equation} 
{\cal L}  _a \supset \frac{\alpha_s}{8\pi } \left( \frac{ a }{  f _a } - \theta \right)    G_{\mu\nu}^A \tilde{G}^{\mu\nu}\,. \label{eq:aGG}
\end{equation} 
To address the strong CP problem, the axion must obtain an expectation value that cancels the $ \theta $-term in the Lagrangian. This occurs dynamically when QCD enters the confining phase, generating an axion potential computed in chiral perturbation theory for two light flavors to be~\cite{Hook:2018dlk},
\begin{equation} 
V ( a ) \simeq - m _\pi ^2 f _\pi ^2 \sqrt{ 1 - \frac{ 4 m _u m _d }{ ( m _{ u } + m _d ) ^2 }\sin ^2 \left( \frac{ a }{ f _a } - \theta \right) }\,.
\label{eq:Va}
\end{equation} 
This potential is minimized when $ a = \theta f _a $, solving the strong CP problem by setting the coefficient in Eq.~\eqref{eq:aGG} to zero. From this point forward, we work with the field expanded around its vacuum expectation value (VEV). The axion mass corresponding to Eq.~\eqref{eq:Va} is given by,
\begin{equation} 
m _a \simeq \frac{ f _\pi m _\pi }{ f _a }\,.
\end{equation} 

We now consider the remaining interactions of the axion field. Since we are primarily interested in dynamics below the weak scale, the phenomenology is largely driven by interactions of the axion to the charged first-generation fermions and photon. Furthermore, we only write the pseudo-scalar interactions and neglect sources of weak violation~\cite{Altmannshofer:2022izm}, leading to the interaction Lagrangian,
\begin{equation}
{\cal L} \supset  \frac{c_u}{2f_a}\partial_\mu a \bar{u} \g^\mu \g^5 u + \frac{c_d}{2f_a}\partial_\mu a \bar{d}\g^\mu \g^5 d + \frac{c_e}{2f_a}\partial_\mu a \bar{e} \g^\mu \g^5 e - \frac{c_{\gamma}}{4 f _a }a F_{\mu\nu}\tilde{F}^{\mu\nu}\,,
\label{eq:axioncouplings}
\end{equation}
with the field strength dual defined as $\tilde{F}^{\mu\nu} \equiv \frac{1}{2}\epsilon^{\mu\nu \alpha\beta}F_{\alpha\beta}$.

In this paper, we consider the two well-known UV completions, the Dine-Fischler-Srednicki-Zhitnitsky
(DFSZ) model~\cite{Dine:1981rt,Zhitnitsky:1980tq} and the Kim-Shifman-Vainshtein-Zakharov (KSVZ)~\cite{Kim:1979if,Shifman:1979if}. In the KSVZ model, a vector-like color-triplet fermion and a complex scalar, both charged under $U(1)_{\text{PQ}}$, are added to the SM Lagrangian. The complex scalar gets a VEV, $ f _a $, spontaneously breaking the $U(1)_{\rm{PQ}}$ and producing a pseudo-Goldstone $a$ (axion). The interactions of the axion with the SM fields are generated at the loop level and are given by\footnote{We do not give expressions for $c_u$ and $c_d$ at loop level in the KSVZ model since they are not relevant to the cosmology and phenomenology of the scenarios we consider in our paper.} 
~\cite{Workman:2022ynf},
\begin{align}
    c_e&  \simeq \frac{3\alpha^2}{4\pi^2}(-1.92)\log \left ( \frac{\Lambda_\chi}{m_e}\right ) \,,  ~c_{\gamma}  = - \frac{\alpha}{2\pi} \left(  1.924 \right) \,, \label{eq:fermion}
\end{align}
where $\alpha$ is the electromagnetic fine-structure constant and $\Lambda_\chi$ is the chiral symmetry-breaking scale. Following Ref.~\cite{Workman:2022ynf}, we take $\Lambda_\chi$ to be $\Lambda _\chi \simeq 1~\text{ GeV}$. Furthermore, the effective coupling of the axion to neutrons and protons is defined as,
\begin{equation} 
{\cal L} \supset \frac{c_n}{2f_a}\partial_\mu a \bar n\gamma^\mu \gamma^5n + \frac{c_p}{2f_a}\partial_\mu a \bar p\gamma^\mu \gamma^5p\,,
\end{equation} 
and is computed by matching to a low-energy effective field theory for the nucleons using inputs from $\beta$-decay measurements and lattice simulations~\cite{di_Cortona_2016}. This leads to,
\begin{equation}
c_n = -0.02\,,~{\rm and}~c_p = - 0.47\,.
\end{equation}

In the DFSZ model, the SM is extended to include a complex weak-singlet scalar field and an additional Higgs doublet. One Higgs doublet ($H_u$) gives mass to the up-type quarks, one Higgs doublet ($H_d$) to down-type quarks. Depending on which Higgs doublet gives mass to the leptons, we refer to the model as DFSZ-I ({$H_d$)} or DFSZ-II ($H_u$). All the scalar fields get VEVs resulting in spontaneous symmetry breaking, and the axion is identified as a linear combination of the corresponding Goldstone bosons. The parameter $\tan \beta$ is the ratio of the up-type to down-type Higgs VEVs. The axion has the quark and photon couplings given by
\begin{equation} 
 c_{\gamma}   =  \frac{\alpha}{2\pi } \left( \frac{ 8 }{ 3 }  -  1.924 \right)  \,, ~~c_u = \frac{1}{3}\cos^2\beta \,,~  {\rm and} ~c_d = \frac{1}{3} \sin^2 \beta.
\end{equation} 
The electron coupling depends on the DFSZ realization with $ c_e = \frac{1}{3}\sin^2 \beta $ for DFSZ-I and $  - \frac{1}{3} \cos ^2 \beta $ for DFSZ-II. The effective proton and neutron couplings are given by
\begin{align} 
 c_p = -0.435 \sin^2 \beta - 0.182 \,~ {\rm and}~c_n = 0.414 \sin^2 \beta - 0.160 \,.
\end{align} 

As our proposed dark matter candidate, we introduce a Dirac fermion, $ \chi $. Both the KSVZ and DFSZ QCD axions can interact with $ \chi $ through the dimension-5 interaction,
\begin{equation} 
{\cal L} \supset \frac{c _\chi  }{ 2f _a } \partial _\mu a \bar{\chi} \gamma ^\mu \gamma _5 \chi \,.
\label{eq:aDM}
\end{equation} 
For convenience, we define a dimensionless coupling $g_{a \chi} \equiv c _\chi m _\chi / f _a $, whose natural value is of order $ m _\chi / f _a $. In this work, we will explore a range of values for the coupling (limited by perturbativity, $  g _{ a \chi } \lesssim 4\pi $). In Secs.~\ref{sec:freezein} and \ref{sec:decoupledfreezein}, we will highlight regimes where QCD axion-mediated dark matter can match the observed relic abundance for natural values of $ g _{ a \chi } $ and still be in line with present experimental constraints. 

\subsection{Experimental Constraints}
\label{sec:constraints}
The QCD axion hypothesis has been extensively tested by a collection of terrestrial and astrophysical experiments (see Ref.~\cite{Workman:2022ynf} for a review). The constraints from meson decays and stellar cooling bounds limit the axion mass to be below the eV scale. In this range, the dominant bounds come from various stellar cooling limits. In this section, we briefly summarize the bounds. 

The strongest bound on the QCD axion-photon coupling is from its influence on the cooling of stars within globular clusters~\cite{Ayala_2014,Dolan:2022kul}. Axion-induced cooling would change the ratios of star types. Consistency between simulations and observations results in the bound, $ \left| c _{ \gamma } \right| / f _a  \lesssim 4.7\times 10^{-11}\text{ GeV}^{-1}$~\cite{Dolan:2022kul}. The dominant bound on the electron coupling is from the non-observation of excess cooling in red giant stars. This gives the constraint, $ \left| c _e \right|  / f _a \lesssim 2.9\times 10^{-10}\text{ GeV}^{-1}$~\cite{Straniero_2020}. Finally, the most stringent bound on the axion-nucleon interaction is from the observation of neutrinos ejected during SN 1987A, which would not be measurable if the axion rapidly cooled the supernova core. This constrains the combination of neutron and proton couplings~\cite{Carenza_2019} as~\footnote{An additional bound can be set directly on the gluon coupling by the requirement that relativistic axions are not overproduced in the early universe in conflict with measurements of the effective number of relativistic degrees of freedom. This puts a milder constraint on the decay constant than those previously mentioned~\cite{Caloni_2022}.}
\begin{equation}
\frac{1}{ f _a }\left( c_n^2  + 0.53~c_pc_n + 0.61~c _p ^2\right) ^{ 1/2}\lesssim  9.7 \times 10 ^{ - 10} ~{\rm GeV} ^{ ^{ - 1}}\,.
\label{eq:SN}
\end{equation}

These bounds can be read in terms of bounds on the KSVZ and DFSZ parameter space. For the KSVZ axion, the strongest bound on the decay constant, $f_a$, is from the axion-nucleon interaction, while for the DFSZ axion, the most stringent bound depends on $ \beta $ but is from the axion-electron coupling for $\beta \gtrsim 0.5$ ($\beta \lesssim 0.95$) for DFSZ-I (DFSZ-II). In summary, in this limit, we find 
\begin{equation} 
f _a \gtrsim \left\{ \begin{array}{ll} 3.9 \times 10 ^{ 8} ~{\rm GeV} & ( {\rm KSVZ} ) \\ 1.2 \times 10 ^{ 9}~{\rm GeV}~\sin^2 \beta& ( \text{DFSZ-I} )  \\ 1.2 \times 10 ^{ 9}~{\rm GeV}~\cos^2 \beta & ( \text{DFSZ-II} )\end{array} \right.\,.
\label{eq:facooling}
\end{equation}

It is interesting that, in addition to the stellar bounds just discussed, there are hints of anomalous cooling in systems whose dominant cooling could come from axion bremsstrahlung off of electrons~\cite{Giannotti:2015dwa,Giannotti:2017hny}. The combination of data on the cooling of these several systems -- white dwarf, red giant branch, and horizontal branch stars --  suggests a best-fit value of $ 1.6 \times 10 ^{ - 10}~{\rm{GeV}}^{-1}\lesssim \left| c_{e} \right|  / f _a  \lesssim 2.9 \times 10 ^{ - 10}~{\rm{GeV}}^{-1} $~\cite{Giannotti:2017hny},
with so far little evidence for any other interaction. For the DFSZ-I axion, this corresponds to a suggested region $ 1.2 \times 10 ^{ 9}\sin^2\beta \lesssim f _a/ {\rm GeV}\lesssim 2.1 \times 10 ^9 \sin^2\beta  $. A similar relation applies for DFSZ-II, but with $ \sin \beta \rightarrow \cos \beta $. There is no preferred region in the KSVZ model since the electron coupling is zero at the tree level, and therefore the value of $f_a$ needed to explain these hints is excluded by SN 1987A bounds.

In addition to these astrophysical bounds, upcoming terrestrial experiments will give complementary information. The upcoming helioscope experiment IAXO will either discover the axion or further constrain the axion-photon coupling at low $ m _a $ to $|c_\gamma|/f_a \lesssim 4.4 \times 10 ^{ - 12}  \text{ GeV}^{-1}$, assuming that solar axions are solely produced from the axion-photon coupling. Production from the axion-electron coupling instead provides a bound on the product of the electron and photon couplings. The projected bound depends on $ m _a $ but can be as stringent as $  \sqrt{|c_e c_\gamma|}/f_a \lesssim  2.2 \times 10 ^{ - 11}$ GeV$^{-1}$ ~\cite{IAXO:2019mpb}. For the KSVZ (DFSZ) axion, the bound on the photon coupling corresponds to $ f _a \gtrsim 4.3 \times 10 ^{ 8} ~{\rm GeV} $ ($ f _a \gtrsim 1.0 \times 10 ^{ 8} ~{\rm GeV} $). For the DFSZ axion, depending on the value of $ \beta $, the bound on the combination of the electron and photon couplings corresponds to $ f _a $ as large as $ 7 \times 10 ^{ 8} ~{\rm GeV} $, while for the KSVZ axion the small value of $  c _e $ indicates this bound is not competitive with the bound exploiting the photon coupling. Collectively, these future bounds represent an order-of-magnitude improvement over the current bound from the CERN Axion Solar Telescope (CAST) experiment \cite{CAST:2004gzq, CAST:2007jps, CAST:2008ixs, CAST:2013bqn}. Therefore, IAXO will be able to offer a much-needed direct complementary probe of the properties of the axion given the uncertainties associated with stellar cooling bounds (in particular, SN 1987A). The ALPS II experiment will also push constraints on $ c _\gamma $, however is not well-suited to probe the QCD axion line ~\cite{Bahre:2013ywa}.

The axion-DM coupling is constrained by the observation that DM is cold and collision-less on galactic scales. Halo dynamics remain relatively unaffected by forward scattering, where the forward momentum is approximately unchanged. This is because forward scattering is inefficient at transporting heat from the outer region to the inner region of the halo. 
To weigh collisions with greater forward momentum loss more strongly (i.e., those collisions which lead to thermalization of the inner halo), a commonly used weighted cross section is the transfer cross section~\cite{Kahlhoefer:2013dca},\footnote{An alternative cross section is the viscosity cross section. However, following Ref.~\cite{Tulin:2013teo}, we use the transfer cross section to be consistent with most of the dark matter literature.}
\begin{equation}
\label{transfercross}
    \sigma_T \equiv  \int d\Omega\,\frac{d\sigma_{\text{SIDM}}}{d\Omega}(1-\cos\theta)\,,
\end{equation}
where $ \sigma _{ {\rm SIDM}} $ is the self-interacting dark matter cross section. DM self-scatters through three processes at tree level ($\chi \chi \to \chi \chi$, $\bar{\chi} \bar{\chi} \to \bar{\chi} \bar{\chi}$, $\chi \bar{\chi} \to \chi \bar{\chi}$) and we use $\sigma_{\text{SIDM}}$ to denote the average of all three processes. Assuming $m_a \ll m_\chi v/2$ the non-relativistic transfer cross section is~\footnote{We do not consider the effects of Sommerfeld enhancement at non-relativistic velocities on the SIDM cross section in the light of Ref.~\cite{Agrawal_2020}, which shows that pseudoscalar mediated processes are subject only to a miniscule enhancement.}

\begin{equation} 
    \sigma_T = \frac{g_{a\chi}^4}{128 \pi m_\chi^2}\,.
\end{equation}
The self-interaction scattering processes are below the experimental bounds for $\sigma_T/m_\chi\lesssim 1~ {\rm cm} ^2 / {\rm g}$~\cite{Spergel:1999mh,Tulin:2013teo}. This gives the rough constraint,
\begin{equation}
g _{a\chi} \lesssim 0.21 \left ( \frac{m_\chi}{1~\text{MeV}}\right )^{\frac{3}{4}}\,.
\label{eq:gSIDM}
\end{equation}  
Note that some level of self-interaction may even be preferred by current experimental data~\cite{Tulin_2018, https://doi.org/10.48550/arxiv.1904.07915} and is roughly around where Eq.~\eqref{eq:gSIDM} saturates.

Finally, the QCD axion is subject to a cosmological bound that depends on the size of $ T _{ {\rm RH}} $ relative to the temperature at which the PQ symmetry is restored, which we approximate with $ f _a $. If $ T _{ {\rm RH} } \lesssim  f _a  $, the symmetry is never restored in the early universe. In this regime, axions are produced due to the misalignment mechanism with an unknown initial angle. If $ T _{ {\rm RH}} \gtrsim f _a  $, axions are produced from the decay of cosmic defects, and their abundance must be calculated numerically~\cite{Gorghetto:2020qws,Buschmann:2021sdq}. For both regimes of $ T _{ {\rm RH}} $, cosmic axions make up a population of cold dark matter after the QCD phase transition. Requiring this population not to overproduce dark matter sets an approximate bound~\cite{GrillidiCortona:2015jxo,Buschmann:2021sdq},
\begin{equation} 
f _a \lesssim \left\{  \begin{array}{cc} 2.0 \times 10 ^{ 11} ~{\rm GeV} & ( T _{ {\rm RH} } \lesssim  f _a  )   \\ 1.4 \times 10 ^{ 11 } ~{\rm GeV}  & ( T _{ {\rm RH} } \gtrsim   f _a   ) \end{array} \right.\,,
\label{eq:faTRH}
\end{equation} 
where, for concreteness, in the case of $ T _{ {\rm RH} }\lesssim  f _a $, we assumed an initial misalignment angle of 2.15, as employed in Ref.~\cite{GrillidiCortona:2015jxo}.  In our work, we focus on the regime where $ T _{ {\rm RH}} \lesssim f _a $ and show summary plots with this constraint. These bounds have a large degree of uncertainty and should only be taken as a rough guideline for the approximate region of interest. The combination of Eqs.~\eqref{eq:facooling} and \eqref{eq:faTRH} give an experimental target range for $ f  _a $.

\section{Scattering Rates and Thermalization}
\label{sec:history}
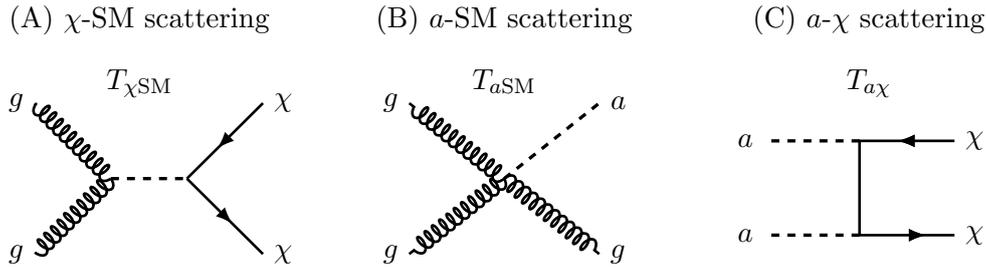
\begin{figure} 
\begin{center} \begin{tikzpicture} 
\coordinate (L) at (0,0);
\coordinate (R) at (12,0);
\coordinate(t3)  at ($(L)!0.1!(R)$);
\coordinate(t2)  at ($(L)!0.5!(R)$);
\coordinate(t1)  at ($(L)!0.9!(R)$);
\node[below] at (t1) {$ T _{a \chi} $};
\node[below]  at(t2) {$ T _{a{\rm SM}} $};
\node[below] at (t3)  {$ T _{\chi{\rm SM}} $} ;
\coordinate(diag1) at ($ (t1)!0.0!(R) +(-0.375,-1.6)$);
\draw[f=0.2,f =0.9 ] ($ (diag1) + (1.5,.5) $) node[right] {$ \chi  $}--++ (-1.25,0) coordinate(S1) --++(0,-1.25) coordinate(S2) --++ (1.25,0) node[right] {$ \chi  $};
\draw[snar](S1) -- ++(-1.25,0)node[left] {$ a $};
\draw[snar](S2) -- ++(-1.25,0) node[left] {$ a $};

\coordinate(diag2) at ($ (t2)!0.0!(R) +(-0,-1.6)$);
\draw[g] (diag2) --++(-1.25,-1.) node[left] {$ g $};
\draw[g] (diag2) --++(-1.25,1.) node[left] {$ g $};
\draw[g] (diag2) --++(1.25,-1)node[right] {$ g $};
\draw[snar] (diag2) --++(1.25,1.) node[right] {$ a $};

\coordinate(diag3) at ($ (t3)!0.0!(R) +(-0.375,-1.6)$);
\draw[g] (diag3) --++(-1,-1.) node[left] {$ g $};
\draw[g] (diag3) --++(-1,1.) node[left] {$ g $};
\draw[snar] (diag3) --++(1,0) coordinate(split);
\draw[f] (split) --++(1,-1)node[right] {$ \chi $};
\draw[fb] (split) --++(1,1) node[right] {$ \chi $};
\node at ($ (t1)+(0,.5) $) {(C)~$ a $-$ \chi $ scattering};
\node at ($ (t2)+(0,.5) $) {(B)~$ a$-SM scattering};
\node at ($ (t3)+(0,.5) $) {(A)~$ \chi $-SM scattering};
  \end{tikzpicture}
\end{center}
\caption{Key processes which can drive dark matter into equilibrium in the early universe.  
The temperatures, $ T _{\chi {\rm SM}} $ and $ T _{ a {\rm SM}} $ denote the temperature at which each process drops out of equilibrium whereas the temperature $T_{a\chi}$ denotes when $aa \leftrightarrow \chi \bar \chi$ comes into thermal equilibrium. For negligible $m_\chi$, the temperature hierarchy depends only on $g_{a\chi}$ and $f_a$ as presented in Eq.~\eqref{eq:heirarchies}.}
\label{fig:temperatures}
\end{figure}

The thermal history of dark matter depends on the interaction strengths of the mediator to the SM ($ \propto 1/f _a $) and the $ \chi $-$ a $ coupling ($ g _{ a \chi } $). There are three scattering processes in the early universe,
\begin{enumerate} 
\item $ \chi $-SM scattering ($\bar{\chi}\chi \leftrightarrow gg$)\,,
\item $ a $-SM scattering ($ag\leftrightarrow \bar{q}q$, $aq \leftrightarrow gq$, $a\bar q \leftrightarrow g\bar q$, $ag \leftrightarrow gg$)\,, and
\item $ a $-$ \chi $ scattering ($\bar{\chi}\chi \leftrightarrow aa$).
\end{enumerate}
We define three temperatures ($ T _{ \chi {\rm SM}} $, $ T _{ a{\rm SM}} $, $ T _{ a\chi} $) corresponding to the epochs at which each rate crosses the Hubble rate, assuming all particles are in thermal equilibrium. Below $T_{\chi \text{SM}}$ $(T_{a\text{SM}})$, the $\chi$-SM ($a$-SM) scattering processes fall out of equilibrium, whereas below $T_{a\chi}$, the $a$-$\chi$ scattering processes come into equilibrium. As we will demonstrate in Sec.~\ref{sec:aSM}, perturbativity of $ g _{ a \chi } $ implies that if $ \chi $ is in thermal contact with the SM, so is the axion (and, therefore, $T_{a\rm{SM}}<T_{\chi\rm{SM}}$). Consequently, there are three possible hierarchies: $ T _{ a \chi } \ll T _{ a {\rm SM}} \ll T _{\chi {\rm SM}} $, $    T _{ a {\rm SM}} \ll T _{ a \chi } \ll T _{\chi {\rm SM}} $, and $   T _{ a {\rm SM}} \ll T _{\chi {\rm SM}}\ll T _{ a \chi } $, depending on the size of the interaction strengths. We will explore dynamics within each hierarchy in this work. A representative Feynman diagram for each scattering process is depicted in Fig.~\ref{fig:temperatures}.

Dark matter can come into thermal contact with the SM directly through $ \chi $-SM scattering or through a combination of $ a$-$\chi $ and $ a $-SM scattering. In such a case, it will undergo freeze-out into SM particles or into axions when it becomes non-relativistic. If $ \chi $ never reaches a thermal abundance, it may still have a sizable relic density due to freeze-in from either SM particles or QCD axion collisions. In either case, calculating the abundance requires knowing the cross sections for the scattering processes above. We now go through each process in turn and calculate the thermalization temperatures to explore the initial conditions for freeze-in or freeze-out.

\subsection{$ \chi $-SM Scattering}
\label{sec:chiSM}
Above the QCD scale, the dominant mechanism through which $ \chi $ can come into thermal equilibrium with the SM is through collisions with gluons and SM fermions. Fig.~\ref{fig:temperatures}~(A) shows the leading diagram which drives the gluon-scattering process. The corresponding cross sections of the $ g g \rightarrow \chi \bar{\chi} $ process and its inverse are,
\begin{align} 
\sigma _{ gg \rightarrow \bar{ \chi} \chi } & = \frac{v _\chi }{8\pi^3}  \frac{g^2_{a\chi} \alpha_s^2}{f_a^2}\,, \label{eq:ggchichi} \\ 
\sigma _{ \bar{ \chi} \chi \rightarrow gg} & = \sigma _{ gg \rightarrow \bar\chi \chi } / v _{ \chi } ^2 \,, \label{eq:chichigg}
\end{align}
where $ v _\chi \equiv \sqrt{1-4 m_\chi^2/ s} $.\footnote{In Eq. \eqref{eq:chichigg}, and everywhere else in this work, we define the cross section to include a sum over the number of internal degrees of freedom of the initial and final states.}

Scattering involving SM fermions can similarly produce and deplete $ \chi $ particles. The corresponding production cross section and its inverse process are
\begin{align}
\sigma_{f\bar{f}\to\chi\bar{\chi}} &= \frac{1}{4\pi s} \frac{v_\chi}{v_f} \left ( \frac{c_f m_f}{f_a} \right )^2 g^2_{a\chi} N_c \, , \label{eq:ffchichi}\\
\sigma_{\chi\bar{\chi}\to f \bar{f}} &= \sigma_{f\bar{f}\to\chi\bar{\chi}} \frac{v_f^2}{v_\chi^2}\,. \label{eq:chichiff}
\end{align}
$N_c$ is the number of colors, and we define $v_f \equiv \sqrt{1-4 m_f^2/s}$. Comparing the cross sections in Eqs.~\eqref{eq:ggchichi} and \eqref{eq:ffchichi}, we observe that the gluon rate will dominate over the fermion rate due to the parametric suppression in the fermion rate $m_f^2/s \sim m_f^2/T^2$. Consequently, the interaction of the DM with the SM fermions is inconsequential for thermalization with the SM bath for generic values of $ T _{ {\rm RH}} $, and we neglect it here. 

To calculate the rate at which the $ g g \rightarrow \chi \bar \chi $ process can drive $ \chi $ into thermal equilibrium with the SM, we compute the scattering rate by carrying out the thermal average over Eq.~\eqref{eq:ggchichi}, assuming a Maxwell-Boltzmann distribution for the gluons and setting $ m _\chi \rightarrow 0 $,\footnote{Throughout this work, we approximate the particle phase space to be Maxwell-Boltzmann and neglect any finite temperature corrections to the particle masses. These assumptions were found to lead to $ {\cal O} ( 1 )  $ corrections to the interaction rates in the case of Sequential freeze-in, where a non-thermalized mediator produces dark matter particles~\cite{Belanger:2020npe}. While we do not consider this case, we note that it is likely $ {\cal O} ( 1 ) $ corrections will also be present in our results.}
\begin{equation}
\label{eq:ratechichigg}
 \Gamma _{ \chi {\rm SM} } \simeq  \frac{1}{4\pi^3}T^3 \left ( \frac{\alpha_s}{8\pi f_a}\right )^2 g^2_{a\chi}\,.
\end{equation} 
Importantly, the $ \chi $-SM scattering is suppressed by both $ g _{ a \chi } ^2 $ and $ 1/ f _a ^2  $. 
We compute the thermalization condition by setting this rate equal to the Hubble rate. We find the corresponding thermalization temperature,
\begin{equation} 
T _{ \chi\text{SM}} \simeq 3.6 \times 10 ^{ 7} ~{\rm GeV} \left( \frac{ f _a }{10 ^{ 9}  ~{\rm GeV}  } \right) ^2 \left( \frac{ 1 }{ g _{a\chi} } \right) ^2 \,.
\label{eq:TchiSM}
\end{equation}

\subsection{$ a $-SM Scattering}
\label{sec:aSM}
The axion-SM conversion process has been extensively studied in the literature. The axion can remain today as a relativistic relic and be probed through its influence on $ \Delta N _{ {\rm eff}} $~\cite{PhysRevD.66.023004,Notari:2022zxo,DEramo:2022nvb} and using experiments designed to search for axion dark matter~\cite{Dror:2021nyr}. Our primary interest here is to determine whether the QCD axion thermalizes with the SM sector. Thermalization occurs primarily through four processes of the same order, $ g g \leftrightarrow g a $, $ g a \leftrightarrow q \bar{q} $, $ a q \leftrightarrow  g q $, and $ a \bar q \leftrightarrow  g \bar q $. A sample Feynman diagram is shown in Fig.~\ref{fig:temperatures}~(B). The full expression for the rate can be found in Ref.~\cite{PhysRevD.66.023004}. We approximate the strong coupling constant with its value at $ 10~{\rm TeV} $, $\alpha _s (10 \text{ TeV} )  \simeq 0.075 $,\footnote{We find an $\mathcal O(1)$ variation in the rate if, instead, we take the strong coupling constant to be $\alpha_s (10^5 \text{ TeV}) \approx 0.042$. More generally, solving for the thermalization temperature numerically, allowing for the running of $\alpha_s$, leads to the same qualitative results, a part from $\mathcal O(1)$ corrections.} and we find the collision rate 
\begin{equation} 
 \Gamma_{a\text{SM}} \simeq 6.8 \times 10^{-5}\frac{T^3}{f_a^2}\,.
\end{equation}

By comparing the $ a-{\rm SM} $ collision rate to the Hubble rate, we estimate the thermalization temperature as, 
\begin{align}
T_{a\text{SM}}&\simeq 2\times 10^4~\text{GeV} \left ( \frac{f_a}{10^9~\text{GeV}}\right )^2  \,.
\label{eq:TaSM}
\end{align}
Note that unlike for $ \chi $-SM scattering, $ a $-SM scattering is not parametrically-suppressed by $ g _{ a \chi } ^2 $, but only $ 1 /f _a ^2 $. Comparing Eqs.~\eqref{eq:TchiSM} and \eqref{eq:TaSM}, we observe that $ T _{ a\text{SM}} \ll T _{ \chi \text{SM}} $ for all $ g _{ a \chi } $ in the perturbative regime, as we expected following the discussion in the introduction of Sec. \ref{sec:history}.

\subsection{$ a $-$ \chi $ Scattering}
If the axion reaches a thermal distribution with the SM, axion-$ \chi $ scattering can drive dark matter in thermal equilibrium as well. A sample Feynman diagram is shown in Fig.~\ref{fig:temperatures}~(C). The cross sections for the process $aa \rightarrow \chi\bar{\chi}$ and its inverse process are given by,
\begin{align} 
\sigma_{a a \rightarrow \chi \bar{\chi} }  & = \frac{g_{a\chi}^4}{2 \pi s}   \left[  \text{tanh}^{-1} v _\chi - v _\chi  \right] \,, \label{eq:aatochichi} \\ 
\sigma_{\bar{\chi}\chi \rightarrow  aa}  & = \sigma_{a a \rightarrow \chi \bar{\chi} } / 2v _\chi ^2\,.\label{eq:chichitoaa}
\end{align} 

Using Eq.~\eqref{eq:aatochichi}, we can compute the thermalization rate of $ \chi $ from axion annihilations, assuming the axion has a thermal distribution. In the limit $  4 m _\chi ^2 \ll s $ we find
\begin{equation} 
\Gamma _{ a \chi } =  3 \frac{g_{a\chi}^4}{K_2(\frac{m_\chi}{T})^2} \frac{\zeta(3)}{16 \pi^3} \frac{T^5}{m_\chi^4} \int_{2m_\chi/T}^\infty dz \, \left [ \log \left ( \frac{T}{m_\chi} z \right ) - 1 \right ] z^2 K_1 (z)\,,
\label{eq:Gamma_achi}
\end{equation} 
where $z \equiv  \sqrt{s}/T$ and $K_1$ ($K_2$) is the first (second) modified Bessel function of the second kind. This equation has a IR divergence as $ m _\chi \rightarrow 0 $, forcing us to keep the mass explicitly in the thermalization rate. Equating Eq.~\eqref{eq:Gamma_achi} to the Hubble rate gives an implicit equation for the thermalization temperature,
\begin{equation} 
T _{ a \chi} \simeq g_{a\chi}^4 M_{\text{pl}} \frac{1}{x^4 K_2 \left ( \frac{x}{2} \right )^2} \frac{3 \zeta(3)}{\pi^3}\left [ \int_x^\infty dz \, K_1(z) (z^2 - x^2) \left ( \log \frac{ 2z }{x}  -1 \right ) \right]\,,
\label{eq:achi}
\end{equation} 
where $x = 2 m_\chi/T_{a\chi}$. This equation can be solved numerically for $T_{a\chi}$. We find a result that is largely insensitive to $ m _\chi $:
\begin{equation}
    T_{a\chi} \sim 1 \text{ TeV} \left ( \frac{g_{a\chi}}{10^{-5}} \right )^4\,. \label{eq:Tachi}
\end{equation}
The value of $ T _{ a \chi } $ changes by $\sim 50\%$ 
when $ m _\chi  $ is varied over the keV-TeV range. If axions are in equilibrium with the SM, they will, in turn, thermalize $ \chi $'s if the temperature of the universe is below $ T _{ a \chi }$.

The parameters $f_a$ and $g_{a\chi}$ dictate the temperature hierarchy. Combining Eqs.~\eqref{eq:TchiSM}, \eqref{eq:TaSM}, and \eqref{eq:Tachi}, we find,
\begin{equation}
\begin{tikzpicture}[baseline=(current  bounding  box.center)]
\node at (-6,0) {$   T_{a\chi} \ll T_{a\text{SM}} \ll T_{\chi\text{SM}} :$};
\node at (-6,-1) {$   T_{a\text{SM}} \ll T_{a\chi} \ll T_{\chi \text{SM}}:$};
\node at (-6,-2) {$    T_{a\text{SM}} \ll T_{\chi\text{SM}}\ll T_{a\chi} :$};
\node at (3.75,0) {$  g_{a\chi} \ll 2 \times 10^{-5} \left ( \frac{f_a}{10^9~\text{GeV}} \right )^{1/2} $};
\node at (1.5,-1) {$    2 \times 10^{-5} \left ( \frac{f_a}{10^9~\text{GeV}} \right )^{1/2} \ll g_{a\chi} \ll 3\times 10^{-3} \left ( \frac{f_a}{10^9\text{ GeV}} \right )^{1/3}$};
\node at (-0.75,-2) {$    3 \times 10^{-3} \left ( \frac{f_a}{10^9~\text{GeV}} \right )^{1/3} \ll g_{a\chi} $};
\end{tikzpicture}\,.
\label{eq:heirarchies}
\end{equation}

\section{Thermally-Coupled Axions}
\label{sec:coupled}
Armed with the expressions for the thermalization of the dark sector, we now study several different viable thermal histories that lead to the observed dark matter abundance. We consider three different regimes. We begin by studying freeze-out into the SM and show that this mechanism is not viable for any perturbative value of $g  _{ a \chi } $. We then go on to consider freeze-in from the SM, finding that the freeze-in production may be either IR or UV-dominated; the value of $ T _{ {\rm RH}}$ fixes the leading freeze-in mechanism. Finally, we consider freeze-out and freeze-in processes happening purely within the dark sector.

\subsection{SM Freeze-out: $\chi \bar \chi \rightarrow   {\rm SM} $}
\label{sec:DMfreezeout}
Conventional dark-matter freeze-out can proceed through $ \chi $ collisions into SM particles. Comparing the cross sections in Eqs. \eqref{eq:chichigg} and \eqref{eq:chichiff}, we conclude that annihilations into gluons will generically dominate the freeze-out process over annihilations into SM fermions. Since the gluon process is driven by the interaction which solves the strong CP problem, the rate can be calculated independently of the KSVZ or DFSZ UV completion. The thermally-averaged cross section can be calculated starting with Eq.~\eqref{eq:chichigg}, and is given in the limit of small DM velocities 
(the relevant limit during freeze-out) by
\begin{equation}
    \langle \sigma_{\chi \bar{\chi} \to gg} v \rangle \simeq \frac{1}{4\pi^3}  \frac{g^2_{a\chi} \alpha_s^2}{f_a^2}\,. \label{eq:sigmav}
\end{equation}
The $ \chi $ Boltzmann equation is,
\begin{equation} 
\dot{ n} _\chi  + 3 H n _\chi = \frac{1}{2} (\bar{n}^2_{\chi} - n _\chi ^2) \left\langle \sigma _{ \chi \bar{\chi} \rightarrow  g g }v \right\rangle \,.\label{eq:BE}
\end{equation} 
where $\bar{n}_{\chi}$ is the number density of $\chi$ in thermal equilibrium.
Integrating Eq.~\eqref{eq:BE}, we find the dark matter yield, $ Y _\chi \equiv n _\chi / s $ 
and the relic density condition,
\begin{equation}
\Omega _\chi \simeq \Omega _{ {\rm DM}} \left( \frac{ 1 }{ g _{ a \chi } } \right) ^2  \left ( \frac{f_a}{2.4\times 10^3~\text{GeV}} \right )^2\,,
\end{equation}
where $ \Omega _{ {\rm DM}} \simeq 0.27 $. Requiring $ g _{ a \chi } \lesssim 4\pi $ to fulfill perturbative unitarity, and requiring $\Omega_\chi=\Omega_{\rm{DM}}$ leads to $ f _a \lesssim 10 ^{ 4 }~\text{GeV} $ that is well excluded by a combination of neutron star cooling and supernova bounds for both KSVZ and DFSZ axion models (see Sec.~\ref{sec:constraints}). 

\subsection{SM Freeze-in: $ {\rm SM} \rightarrow \chi \bar \chi $}
\label{sec:freezein}
Existing experimental bounds on the value of $f _a $ suggest that dark matter may freeze-in through the QCD axion portal. This case corresponds to the case where $ T_{\text{RH}} \ll T_{\chi \text{SM}} $ such that dark matter is not thermalized. For simplicity, we also assume $ T _{ {\rm RH}} \ll T _{ a {\rm SM}} $ such that the axion is not significantly populated in the early universe. This is not strictly necessary, but simplifies the analysis since, in this case, $ \chi $'s are not additionally significantly produced from axion annihilations.

The parametric dependence of the relic abundance condition depends on the dominant production process. At high temperatures, the non-renormalizable nature of the gluon interaction makes the $ gg \rightarrow \chi \bar \chi $ collision rate sensitive to a positive power of temperature (see Eq.~\eqref{eq:ratechichigg}). Hence the structure of the interaction results in a freeze-in abundance scaling as a positive power of the reheating temperature of the universe, a situation known as ``UV freeze-in.'' 

Remarkably, abundances produced via axion-fermion interactions are largely insensitive to $ T _{ {\rm RH}} $.  This might be puzzling since the axion interactions arise as dimension-5 operators, violating the intuition laid out in Ref.~\cite{Hall:2009bx}. However, in the case of elementary external states, the amplitudes pick up a power of the fermion mass, reducing the UV sensitivity. The fermion mass insertion makes heavier fermions an important contribution, if the fermion is not too heavy such that its density is significantly Boltzmann suppressed at $ T _{ {\rm RH}} $. As we will see, top anti-top annihilations can be a significant source of dark matter for a range of $ T _{ {\rm RH}} $. 

For composite external states (e.g., SM pions), the amplitude picks up contributions of order the QCD scale. This feature can result in a substantial freeze-in abundance from QCD resonances for $T_{\text{RH}} \lesssim 100~{\rm GeV}$. The abundance in this case is also insensitive to $ T _{ {\rm RH}} $, resulting in a ``IR freeze-in''.

\begin{figure}[t!]
\centering
\includegraphics[width=12cm]{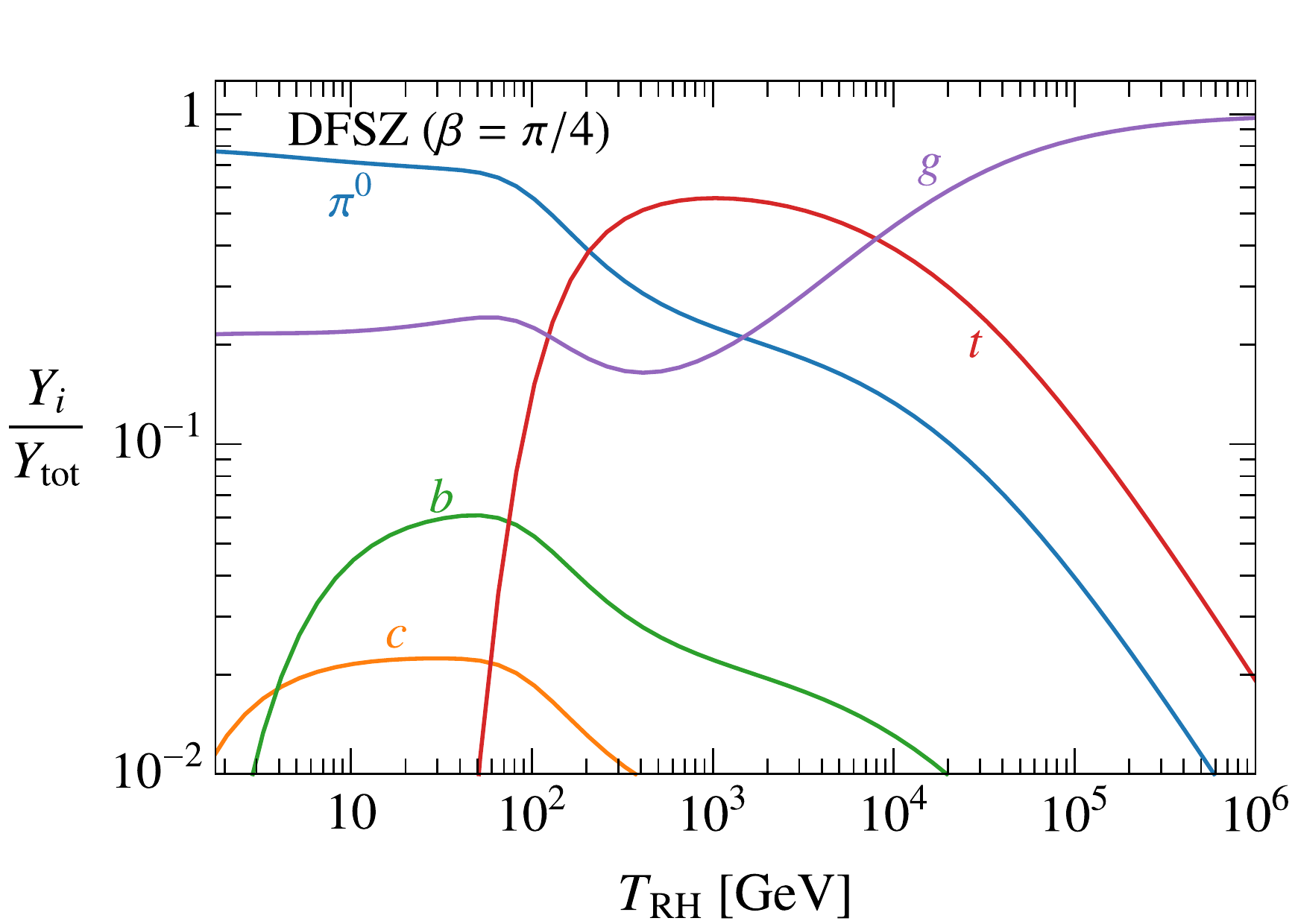}
\caption{The ratio of the dark matter yield from individual processes to its total yield from all processes for the DFSZ model with $ \beta = \pi / 4 $. We show curves from $\pi^0 \to \bar{\chi}\chi$ ({\bf \color{c1} blue}), $gg \to \bar{\chi}\chi$ ({\bf \color{c5} purple}), $t\bar{t} \to \bar{\chi} \chi$ ({\bf \color{c4} red}), $b\bar{b}\to \bar{\chi}\chi$ ({\bf \color{c3} green}) and $c \bar{c} \to \bar{\chi}\chi$ ({\bf \color{c2} orange}), as a function of the reheating temperature of the universe in the limit where $ m _\chi \rightarrow 0 $. The dependence of $ g _{ a \chi } $ and $ f _a $ drops out in calculating the ratio. We conclude that for $ T _{ {\rm RH} }$ below $10^4$ GeV, IR-dominated top and pion contributions can be sizable, whereas above $10^4$ GeV, the UV-sensitive gluon contribution is the dominant one.} 
\label{fig:ratio}
\end{figure}

In the following sections, we explore these different processes. The most important contributions are from gluon annihilations, top anti-top annihilations, and pion decay. Which process dominates will depend on $ T _{ {\rm RH}} $, and we show the ratios of the different process yields (for which the $ f _a $ and $ g _{ a \chi } $ dependence cancels), assuming $ m _\chi $ is small in Fig.~\ref{fig:ratio}. For concreteness, we produce the plot for the DFSZ model with $ \beta = \pi /4 $.

In all of the cases discussed above, we can have additional dynamics if $g_{a\chi}$ is sufficiently large to cause the process $\chi\bar{\chi} \to a a$ to come into local thermal equilibrium. In this case, $ \chi $'s will annihilate after production and calculating the abundance requires tracking the evolution of the $ \chi $ phase space. While it's possible for $ \chi $ to have the observed relic abundance in this regime, we leave this more involved calculation for future work.

\subsubsection{$ g g \rightarrow \chi \bar{\chi} $}
\label{sec:gluon}
The freeze-in abundance from gluon annihilations is UV-dominated due to the higher-dimensional nature of the gluon interaction. The Boltzmann equation for the $ \chi $ number density is given by, 
\begin{equation}
\label{eq:Bol}
\dot{n}_\chi + 3 H n_\chi = 2 \bar{n}_{g} ^2 \left\langle \sigma _{ g g \rightarrow \chi \bar{\chi} } v \right\rangle \,,
\end{equation}
where $\bar{n}_{g}$ is the gluon number density in thermal equilibrium. The collision term depends on the gluon phase space density, which we approximate as a Maxwell-Boltzmann distribution during freeze-in. Assuming that the DM mass $m_\chi$ is negligible compared to the typical center-of-mass energy of the incoming gluons (as expected from a UV sensitive process), we find the collision term, 
\begin{equation}
\bar{n}_{g} ^2 \left\langle \sigma _{ g g \rightarrow \chi \bar{\chi} } v \right\rangle  = \frac{4}{\pi^5} \left ( \frac{\alpha_s}{8\pi f_a} \right )^2 g^2_{a\chi} T^6\,.
\end{equation}

Integrating the Boltzmann equation from reheating until today gives the yield, 
\begin{equation}
\label{eq:gluondriven}
Y_{\chi} \simeq \frac{1}{ 8 \pi ^7 }\left( \frac{g _{ a \chi } }{ f _a } \right) ^2  \frac{T^6_{\text{RH}}}{s_{\text{SM}}(T_{\text{RH}}) H(T_{\text{RH}})} \int \displaylimits_{a_{\text{RH}}/a_{\text{QCD}}}^1 \frac{\alpha_s^2 }{g_{*,s} \sqrt{g_*}} \, dz \,,
\end{equation}
where we use $ Y _\chi $ to denote the sum of the particle and anti-particle yields, $z \equiv a_{\text{RH}}/a$, and $s_{\text{SM}}$ is the entropy density of the SM bath. Evaluating this integral numerically gives the relic density condition, 
\begin{equation}
\Omega _\chi \simeq  \Omega _{ {\rm DM}}  \bigg (\frac{10^9~\text{GeV}}{f_a} \bigg )^4  \bigg (\frac{m_\chi}{370~\text{GeV}}\bigg )^3 \bigg ( \frac{g_{a\chi}}{m_\chi/f_a} \bigg )^2  \bigg ( \frac{T_{\text{RH}}}{10^{10}~\text{GeV}}\bigg )\,. \label{eq:ggrelicdensity}
\end{equation}
We conclude that for weak-scale dark matter and high reheating temperature, the QCD axion can successfully mediate dark matter freeze-in, even for natural values of $ g _{ a \chi } $. The UV nature of this process makes it the dominant freeze-in process for $T_{\rm{RH}}\gtrsim 10^4$ GeV, as seen in Fig.~\ref{fig:ratio}.

\begin{figure}[t!]
\centering
\includegraphics[width=7.5cm]{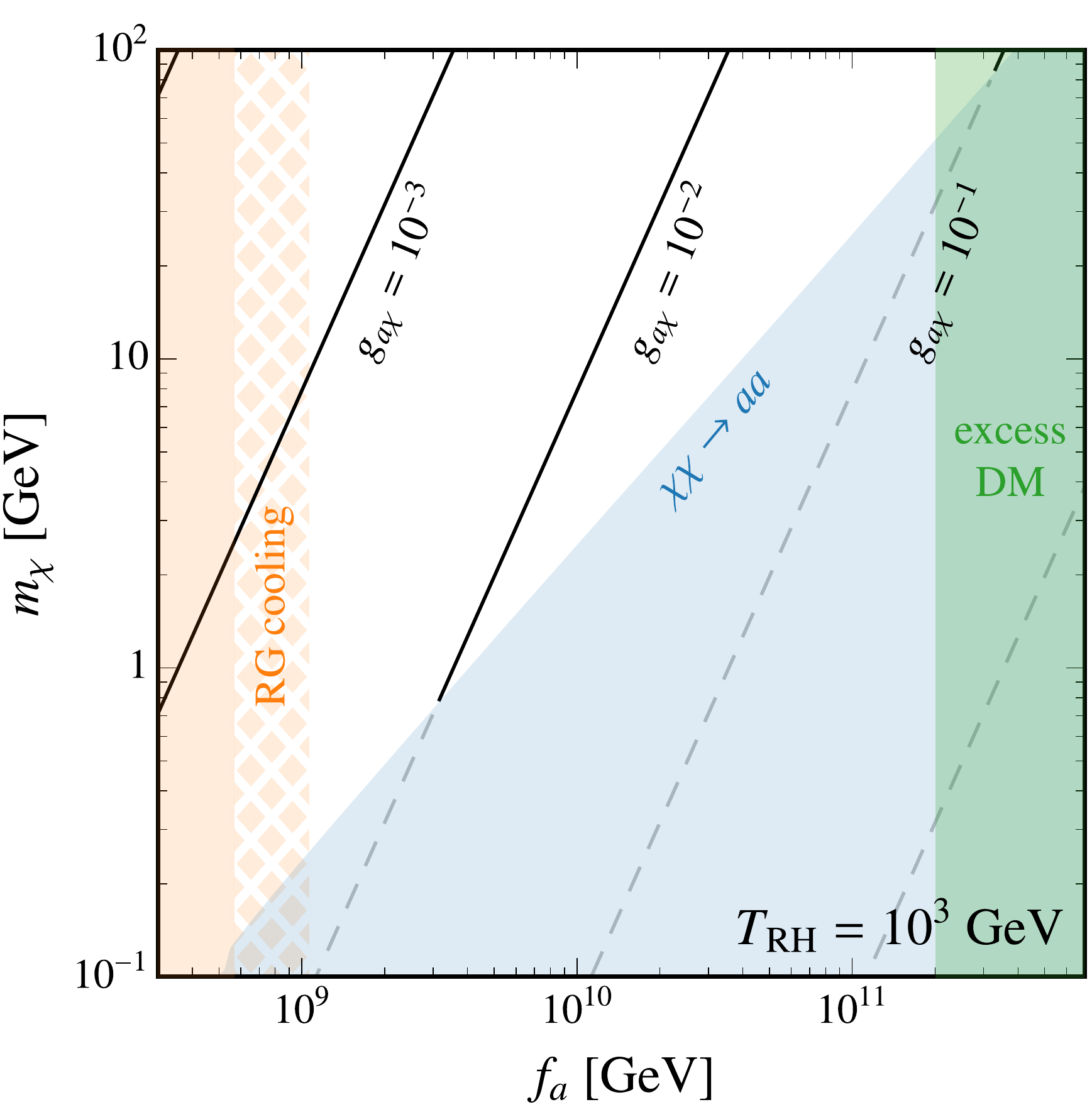}
\includegraphics[width=7.5cm]{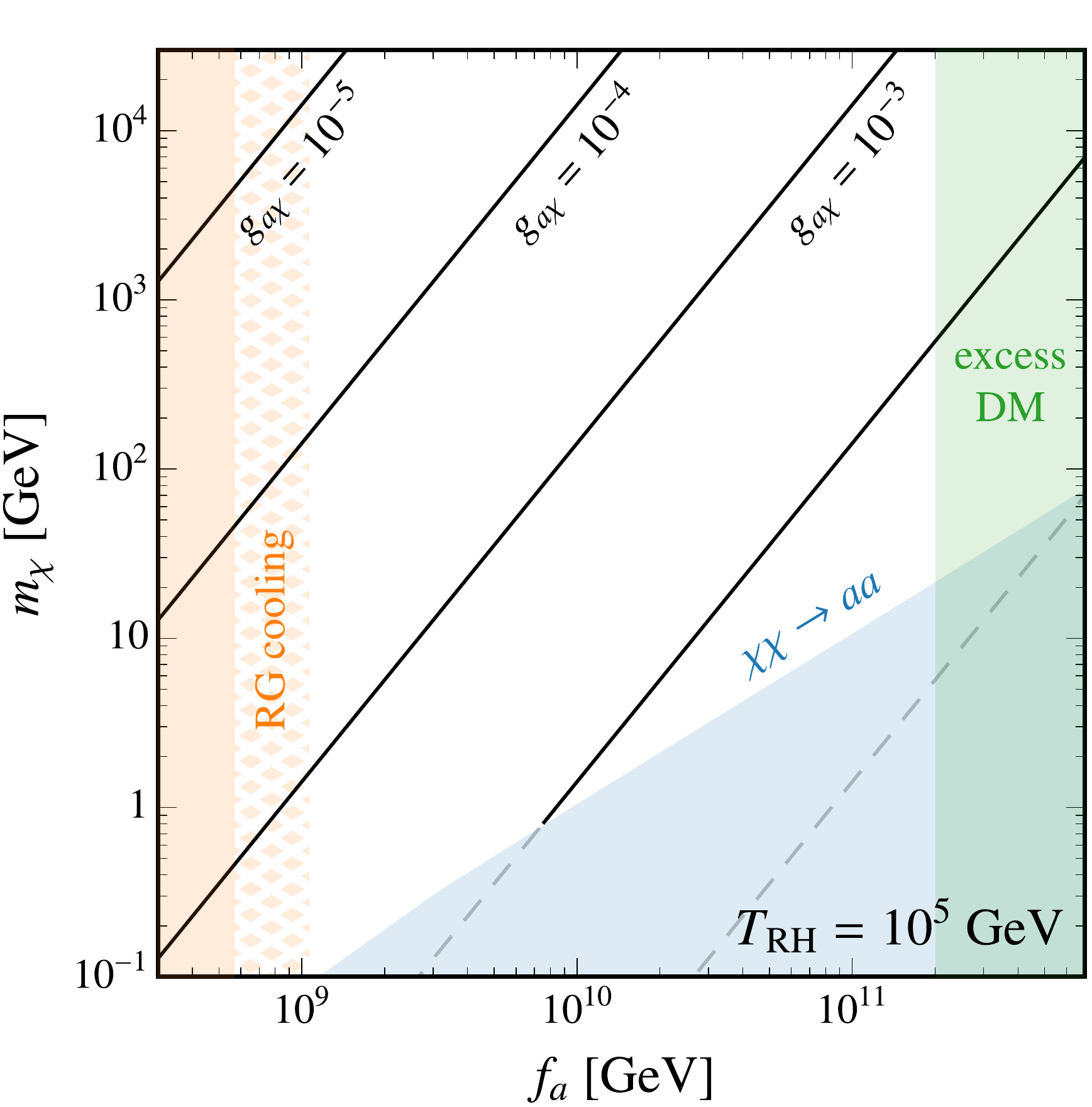}
\caption{The parameter space of dark matter frozen in through the Standard Model-axion interactions for $ T _{ {\rm RH}} = 10 ^3~{\rm GeV} $ ({\bf Left}) and $ 10 ^5~{\rm GeV} $ ({\bf Right}). For concreteness, we focus on the DFSZ axion with $\beta=\pi/4$ which has sizable contributions from $ t \bar{t} \rightarrow \chi \chi $ and $ g g \rightarrow \chi \chi $, depending on $ T _{ {\rm RH}} $ (see Fig.~\ref{fig:ratio}). We show the strongest astrophysical bound on the axion from red giant cooling ({\bf \color{c2} orange}), depicting the region corresponding to the cooling hints with cross-hatches. We also overlay the bound from excessive dark matter production ({\bf \color{c3} green}) and the region in which DM can subsequently freeze-out by annihilating into axions (\color{c1} blue).} 
\label{fig:mchi_fchi}
\end{figure}

We show the parameter space for which QCD axion-mediated dark matter obtains the measured relic density from gluon freeze-in for $ T _{ {\rm RH}} = 10 ^{ 5}~{\rm GeV} $ as black lines in Fig.~\ref{fig:mchi_fchi} ({\bf right}). We show curves for $ g _{ a \chi } = 10  ^{ - 5} $, $ 10 ^{ - 4} $, and $ 10 ^{ - 3} $. For contrast, we show the constraints from excessive red giant cooling in solid orange (the preferred region is shown in cross-hatch) and excessive dark matter production from misalignment in green (see Sec. \ref{sec:constraints}). 

We now comment on the regime of validity of this scenario. If $ g _{   a \chi } $ is large, then it is possible for the dark matter to subsequently freeze-out by annihilating into axions, $\bar \chi\chi\to aa$. A proper investigation of the relic abundance condition in this case requires tracking the evolution of the phase space density of the dark matter by solving the unintegrated Boltzmann equations. Since we do not calculate these effects in this work, we simply check if $n_\chi \sigma _{ \chi \bar \chi \rightarrow a a } \lesssim   H$, at any point after dark matter is produced.\footnote{Since the dark matter annihilation cross section scales as inversely with energy squared, this condition is the most stringent when dark matter is about to become non-relativistic. To estimate this condition, we fix the $ \chi $ kinetic energy of order $ m _\chi $ and its number density to $Y _\chi s _{ {\rm SM}} $.} This condition is violated in the blue region shown in Fig.~\ref{fig:mchi_fchi}.\footnote{We calculate the relic abundance in the extreme limit where $ \chi $ collisions are rapid enough to thermalize $\chi $ in Sec.~\ref{sec:decoupled}.} Furthermore, to derive the relic abundance, we assumed that $ \chi $ is not produced through other means. In particular, if the axion is in thermal contact with the SM, then it will have a large number density in the early universe and its annihilations may produce dark matter as well. This will not be the case as long as $ T _{ {\rm RH}} \lesssim T _{ a {\rm SM} } $, a condition which is satisfied in the bulk of the parameter space shown in Fig.\ref{fig:mchi_fchi} ({\bf right}).
 
\subsubsection{$ f \bar{f} \rightarrow \chi \bar{\chi} $}
\label{sec:top}
Dark matter can be frozen-in through collisions of SM fermions, $ f $, i.e. by the process $f \bar{f }  \to \chi  \bar{\chi}$. As we will show, the abundance is sensitive to the fermion mass. Consequently, for reheating temperatures near the weak scale, the dominant source of dark matter will be from top anti-top annihilations.

The Boltzmann equation can be written as
\begin{equation}
\dot{n}_\chi + 3 H n_\chi = 2\bar{n}_f  ^2 \left\langle \sigma _{  f \bar{f}  \rightarrow  \chi  \bar{\chi}} v \right\rangle\,,
\end{equation}
where $\bar{n}_{f}$ is the fermion number density in thermal equilibrium. In the limit that dark matter is light compared to the fermion $m_\chi^2 \ll m_f^2$,\footnote{In the opposite limit, $m_f \ll m_\chi$, the IR freeze-in abundance is suppressed by an additional factor of $m_f/m_\chi$ such that this is not an efficient production mechanism.} we find that the collision term is,
\begin{equation}
    \bar{n}_f  ^2 \left\langle \sigma _{ f \bar{f} \rightarrow \chi \bar{\chi} } v \right\rangle  = \frac{1}{32\pi^5} N_c \left (\frac{c_f m_f}{f_a} \right)^2 g^2_{a\chi} m_f^2 T^2 K_1 \left ( \frac{m_f}{T} \right )^2 \,.
\end{equation} 

Integrating the Boltzmann equation gives the dark matter yield,
\begin{equation}
    Y_{\chi} \simeq \frac{45}{\pi^2 (1.66)} \frac{1}{32 \pi^5} N_c c^2_{f} g^2_{a\chi} \frac{M_{\text{pl}} m_f}{f_a^2} \int^{\infty }_{m _f / T_{\text{RH}}} dx \, \frac{1}{g_{\star,S}\sqrt{g_{\star}} }x ^{2}  K_1 ( x ) ^2 \,, \label{eq:Yff}
\end{equation}
where $ x \equiv m _f / T $. Contrary to the yield obtained from gluons in Eq.~\eqref{eq:gluondriven}, fermion annihilation into $ \chi $ is approximately $ T _{ {\rm RH}} $ independent, making it an IR-dominated process. We show the abundance from the top quark, bottom quark, and charm quark annihilations relative to the total abundance in Fig.~\ref{fig:ratio}. The factor of $ m _f $ in Eq.~\eqref{eq:Yff} shows that the heavier fermions produce dark matter most efficiently (assuming that $ T _{ {\rm RH}} \gg m _f $). This drives the top quark to contribute the most to the dark matter relic density. For the top quark, the dark matter relic abundance is given by
\begin{equation}
\label{eq:topFI}
\Omega _{ \chi } \simeq  \Omega _{ {\rm DM}} \bigg ( \frac{10^9~\text{GeV}}{f_a}\bigg )^2 \bigg (  \frac{m_\chi}{1~\text{GeV}}\bigg ) \bigg ( \frac{g_{a\chi}}{1.4\times 10^{-3}}\bigg )^2 \cos^4 \beta\,.
\end{equation}

The $ \chi $ produced from top annihilations can make up all of dark matter for perturbative values of $ g _{ a \chi } $ (see left panel of Fig. \ref{fig:mchi_fchi}).

\subsubsection{$\pi^0\rightarrow \chi \bar{\chi} $}
\label{sec:pidecay}
Decay of light mesons can also produce dark matter. Since rates involving external composite states are proportional to the QCD scale, these processes are also IR-dominated. As such, we focus on the lightest meson that can produce dark matter - the $ \pi ^0 $. Heavier mesons, such as the $ \rho  $, $ J / \psi $, $ \hdots $ will have similar contributions but the resulting abundance will be suppressed relative to the $ \pi ^0 $ contribution by ratios of heavy meson mass to the $ \pi ^0 $ mass. 

The $ \pi ^0 $ interaction with dark matter arises from the axion-$ \pi ^0 $ mixing. This mixing has been extensively studied in the literature (see, e.g., Ref.~\cite{Bauer:2017ris}). 
Starting with the quark-level Lagrangian in Eq.~\eqref{eq:axioncouplings}, the resulting $ \pi ^0 $-axion chiral Lagrangian is given by,
\begin{equation} 
{\cal L} \supset \frac{1}{2} ( \partial ^\mu a ) ^2 + \frac{1}{2} ( \partial ^\mu \pi ^0 ) ^2  -  \varepsilon \partial ^\mu a \partial _\mu \pi ^0 \,,
\label{eq:Lmixing}
\end{equation} 
with the kinetic mixing coefficient, $\varepsilon$, given by,
\begin{align} \nonumber
 \varepsilon &  \equiv  \frac{f_\pi \sqrt{2}}{4f_a} \left ( c_u - c_d - \frac{(m_d-m_u)}{m_d+m_u} \right )\,, \\ 
& \simeq - 10 ^{ - 11}\left( \frac{ 10 ^{ 9} ~{\rm GeV} }{ f _a } \right) \left\{ \begin{array}{cc} 1.56 &  ( {\rm KSVZ} )  \\ 1.56-1.53 \cos 2 \beta  &  ( {\rm DFSZ} ) \end{array} \right.\,.
\end{align} 
Here we introduced the pion decay constant $ f _\pi \simeq 130 ~{\rm MeV}$. 

Rotating away the kinetic mixing in Eq.~\eqref{eq:Lmixing} results in a $ \pi ^0$-dark matter  interaction from Eq. \eqref{eq:aDM}. This interaction is given by
\begin{equation}
{\cal L} _{\pi } = \frac{ g_{a \chi}  }{ 2 m _\chi  }\varepsilon \partial _\mu \pi ^0  \bar{\chi}\gamma ^\mu \gamma _5 \chi\,,
\end{equation}
where we used that $m_a^2 \ll m_\pi^2$. This term drives the $ \pi ^0 \rightarrow \chi \chi $ freeze-in process. 
To compute the dark matter yield we solve the Boltzmann equation for $ \chi $, 
\begin{equation}
    \dot{n}_{\chi} + 3 H n_\chi = 2\bar{n}_{ \pi ^0 } \langle \Gamma _{ \pi ^0 \rightarrow \chi \bar{ \chi} } \rangle \,, 
\label{eq:boltzmann}
\end{equation}
where,
\begin{equation} 
\bar{n}_{ \pi ^0 } \left\langle \Gamma _{ \pi ^0 \rightarrow \chi \bar{ \chi} } \right\rangle \equiv  \int\, \frac{d^3 p_\pi}{(2\pi)^3} \frac{ m _\pi }{ E _\pi } \varepsilon ^2  v _\chi ^3\frac{ g _{a\chi} ^2 m_\pi^3 }{ 8\pi }f _{ \pi ^0 }\,.
\label{eq:picollision}
\end{equation} 
In Eq.~\eqref{eq:picollision}, we introduced $f_{\pi^0}$ as the pion phase space, which we approximate with a Maxwell-Boltzmann distribution, and we also introduced, $ v _\chi \equiv ( 1 - 4 m _\chi ^2 / m _\pi ^2 ) ^{1/2}  $. Carrying out the integral in Eq.~\eqref{eq:picollision} gives the collision term,
\begin{align}
 \bar{n}_{ \pi ^0 } \left\langle \Gamma _{ \pi ^0 \rightarrow \chi \bar{ \chi} } \right\rangle & \simeq \varepsilon^2 v _\chi ^3 \frac{ g^2_{a\chi}m_\pi^3}{16\pi^3}  TK_1\left (\frac{m_\pi}{T}\right )\,.
\end{align}

In the limit where $ g _{ \star } \simeq g_{\star}(m_\pi) $ and $g_{\star,S} \simeq g_{\star,S}(m_\pi)$ do not change considerably during freeze-in, we find an analytic solution to the Boltzmann equation,
\begin{align}\nonumber
m_\chi Y_\chi &\simeq m_\chi \frac{3 \varepsilon^2 v_\chi^3 g^2_{a\chi}m_\pi^4}{16 \pi^2 s_{\text{SM}}(m_\pi) H(m_\pi)}\,, \\
& \simeq 0.44 \text{ eV} \left ( \frac{\varepsilon}{10^{-11}} \right )^2 \left ( \frac{m_\chi}{10~\text{MeV}} \right ) \left ( \frac{g_{a\chi}}{0.1} \right )^2\,.
\end{align}
We observe that since $ |\varepsilon| \lesssim 10 ^{ - 11} $ and $ m _\chi \lesssim m _\pi / 2 $, for neutral pion decay to produce the observed dark matter abundance requires large values of $ g _{ a \chi } $. For such large values of the dark matter-axion coupling, $ \chi $'s  produced from pion decays will inevitably annihilate significantly after production, thermalizing the dark sector. To solve this system, one must study the unintegrated Boltzmann equations, and we leave such a detailed study for future work.

\subsection{Dark Sector Freeze-out: $ \chi \bar \chi \rightarrow a a $}
\label{sec:decoupled}
Dark matter can enter thermal equilibrium in the early universe if $ T _{ {\rm RH}} \gtrsim T _{ \chi {\rm SM}}  $ or $ T _{ a \chi } \gtrsim  T _{ {\rm RH}} \gtrsim  T _{ a {\rm SM}} $. If $ \chi $ reaches a thermal distribution in the early universe, it may freeze-out into the QCD axion, $\chi \bar{ \chi} \to aa$. In this section, we study this possibility, finding that dark sector freeze-out is viable for perturbative values of $ g _{ a \chi } $. Furthermore, the corresponding QCD axion abundance can be tested by future measurements of $ \Delta N _{ {\rm eff}} $. 

We consider the case where the dark sector decouples from the SM bath at a temperature $T_{a {\rm SM}}$, which we take to be before freeze-out ($ T _{a {\rm SM}} \gg m _\chi $). The changes in the entropy of the SM will result in a dark sector temperature ($ T ' $) being different from $ T $ by a ratio of $ g _{ \star } $ values, which we can compute using conservation of entropy:
\begin{equation}
T' = \left (\frac{g_{\star,S}(T)}{g_{\star,S}(T_{a {\rm SM}})}\right)^{1/3} T\,.
\label{eq:Tprime}
\end{equation}

The Boltzmann equation for freeze-out is given by
\begin{equation}
    \dot{n}_\chi + 3 H n_\chi = -\frac{1}{2} \langle \sigma _{ \chi \bar\chi \rightarrow a a }v \rangle  (n_\chi^2 - \bar{n}_{\chi}^{2}(T'))\,,
\label{eq:chichiaa}
\end{equation}
where the thermally-averaged cross section is,
\begin{equation}
\langle \sigma _{ \chi \bar\chi \rightarrow a  a} v \rangle \simeq \frac{g^4_{a\chi}}{16 \pi m_\chi^2} \frac{ T ' }{ m _\chi } \,.
\end{equation}
Using the sudden freeze-out approximation we find an approximate analytic solution to Eq.~\eqref{eq:chichiaa}:
\begin{equation}
    Y_{\chi} \simeq \frac{4m_\chi}{T_f} \frac{1}{s_{\text{SM}}(m_\chi)} \frac{H(m_\chi)}{\langle \sigma_{\chi \bar{\chi} \to aa} v \rangle (T_f )} \,.
\end{equation}
The relic abundance condition gives, 
\begin{equation}
 \Omega _{ \chi } \simeq \Omega _{ {\rm DM}} \left [ \frac{g_{\star,S}(T_{a {\rm SM}})}{g_{\star,S}(T_f)} \right ]^{1/3}    ~\bigg ( \frac{m_\chi/T_f}{10} \bigg )^2 \bigg ( \frac{g_\star(T_f)}{15} \bigg )^{\frac{1}{2}} \bigg ( \frac{15}{g_{\star ,S}(T_f)} \bigg ) \bigg (\frac{m_\chi}{1~\text{GeV}} \bigg )^2 \bigg ( \frac{4.4 \times 10^{-2}}{g_{a\chi}} \bigg )^4\,.
\end{equation}
The factor in the square brackets is absent when freeze-out occurs before the dark sector decouples from the SM, i.e. $T_{\text{a\text{SM}}} \lesssim m_\chi$.
Note that due to the large value of $ g _{ a \chi } $ necessary to get the observed abundance, freeze-out into QCD axions can only occur for unnaturally large values of $ g _{ a \chi }$ ($ g_{ a \chi }\equiv c_\chi m_\chi/f_a\gg m_\chi/f_a$). 

The $ \chi \bar\chi \rightarrow a a $ freeze-out process does not require any direct coupling between the SM and the dark sector, and hence is challenging to test experimentally. The only direct constraint comes from searches for dark matter self-interactions. Inputting Eq.~\eqref{eq:gSIDM} into the relic abundance condition, we conclude that this does not meaningfully constrain the parameter space for dark matter which obeys the Tremaine-Gunn bound, $ m _\chi \gtrsim 1~{\rm keV} $~\cite{Tremaine:1979we,Boyarsky:2008ju}. Nevertheless, since dark matter annihilates into QCD axions, which remain relativistic today, the theory predicts a significant source of dark radiation. This extra source of energy density in the early universe would modify the predictions for ratios of light nuclei produced during Big Bang Nucleosynthesis (BBN) and the cosmic microwave background (CMB). Currently, BBN limits $ \Delta N _{ {\rm eff}} < 0.4$~\cite{Cyburt:2015mya}, while the CMB constrains $ \Delta N _{ {\rm eff}} < 0.3 $~\cite{Planck:2018vyg}. Upcoming measurements of the CMB are expected to either discover dark radiation or push these bounds by an order of magnitude in the near future~\cite{SPT-3G:2019sok,SimonsObservatory:2018koc,Abazajian:2019eic}. We now study the implications of these measurements on dark sector freeze-out.

First, we note that if $m _\chi \ll 100~{\rm keV} $, then dark matter would be relativistic at BBN. This would amount to a value of $ \Delta N _{ {\rm eff}} $ in contrast with current limits. Therefore, we assume $  m _{ \chi } \gg 100~{\rm keV} $ such that the relativistic degrees of freedom in the dark sector at and beyond BBN is only given by the axion. We find the value of $ \Delta N _{{\rm eff}} $ predicted by freeze-out of $ \chi $ into QCD axions by tracking the entropy changes in the dark sector. Since the number of effective relativistic degrees of freedom decreases in the SM after $T_{a {\rm SM}}$, the SM photon bath will be warmer than the dark sector.

We are interested in the temperature of the dark sector at a SM temperature, $ T $, well after freeze-out (during either BBN or recombination). At early times, both $ \chi $ and $ a $ are relativistic while after freeze-out only the axion remains relativistic. Using conservation of entropy, we find,
\begin{equation}
T^{\prime } = \left( \frac{  ( 7/8 ) g _\chi +g _a }{ g _a }   \frac{g_{\star,S}(T)}{g_{\star,S}(T_{a {\rm SM}})} \right) ^{1/3} T\,, 
\label{eq:TprimeFO}
  \end{equation}
where $ g _a = 1 $ and $ g _\chi = 4  $ count the number of degrees of freedom in the dark sector. The effective number of relativistic degrees of freedom in addition to the photons and neutrinos is given by,
\begin{align}\nonumber
    \Delta N_{\rm eff} & = \frac{1}{2} \frac{8  }{7}  \left( \frac{ 11 }{ 4 } \right) ^{ 4/3} g _a \left( \frac{ ( 7/8 )  g _\chi + g _a }{ g _a } \frac{ g _{ \star, S } ( T ) }{ g _{ \star, S } ( T _{a {\rm SM}} ) } \right) ^{ 4/3}\,, \\ 
& \simeq  \left( \frac{ 8.1 \times g _{ \star,S } ( T ) }{  g _{ \star ,S} ( T _{a {\rm SM}} )} \right) ^{ 4/3}\,. 
\end{align}
While in this section, we have treated $ T _{ a {\rm SM}} $ as a free parameter, its value is related to $ f _a $ through Eq.~\eqref{eq:TaSM}. Consequently, existing constraints on $ f _a $ limit its value to well above the weak scale. Nevertheless, we consider it as a free parameter given that it may be modified by introducing additional particles to the spectrum.

\begin{figure}[t!]
\centering
\includegraphics[width=12cm]{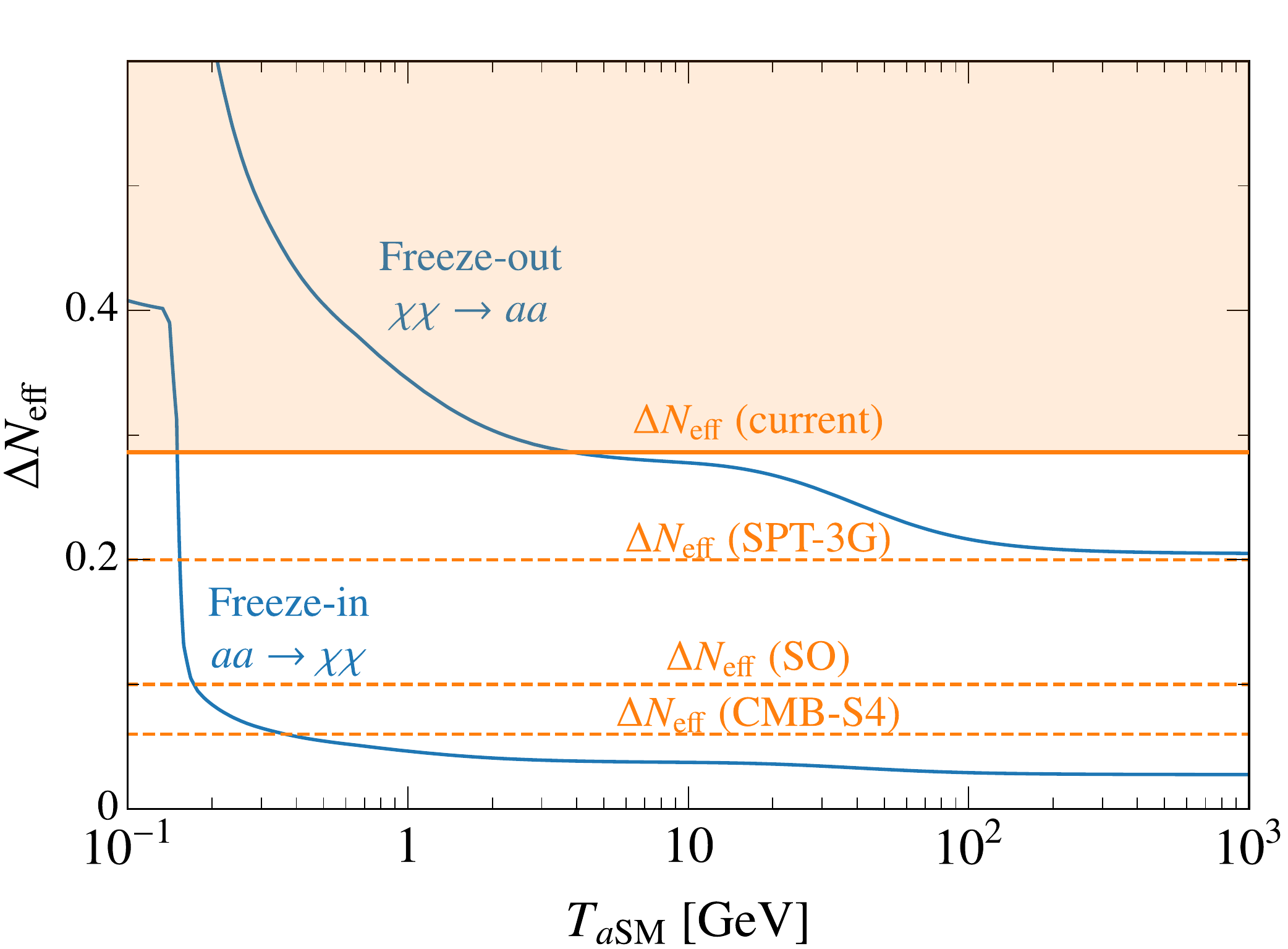}
\caption{The constraints on $ \chi \bar\chi \rightarrow a a $ freeze-out from the non-observation of dark radiation. The constraint is shown as a function of the temperature at which relativistic decoupling occurs between the dark and visible sector ($ T _{a {\rm SM}} $). Current (Planck collaboration~\cite{Planck:2018vyg}) and future (SPT-3G~\cite{SPT-3G:2019sok}, the Simon's Observatory (SO)~\cite{SimonsObservatory:2018koc}, and CMB-S$ 4 $~\cite{Abazajian:2019eic}) constraints on $\Delta N_{\rm{eff}}$ at the $2\sigma$ level are shown.}
\label{fig:Dark_Rad_Ti}
\end{figure}

The value of $ \Delta N _{ {\rm eff}} $, as evaluated well after electron decoupling, is plotted in Fig.~\ref{fig:Dark_Rad_Ti} as a function of $ T _{a {\rm SM}} $. We see that if the axion-Standard Model processes decouple before temperatures of $ 3.7 ~{\rm GeV} $, then the value of $ \Delta  N _{ {\rm eff}} $ is in conflict with current measurements. Furthermore, future SPT-3G~\cite{SPT-3G:2019sok}, Simon's Observatory~\cite{SimonsObservatory:2018koc}, and CMB-S4~\cite{Abazajian:2019eic} observations will either discover a non-zero $ \Delta N _{ {\rm eff}} $ or rule out the possibility of $ \chi \bar \chi \rightarrow a a $ freeze-out entirely, even for large values of $T_{a{\rm{SM}}}$ as computed in Eq. (\ref{eq:TaSM}).

\subsection{Dark Sector Freeze-in: $ a a \rightarrow \chi \bar\chi $}
\label{sec:decoupledfreezein}
If dark matter never comes into thermal contact with the Standard Model bath, then it may still be frozen-in through axion collisions, $  a a \rightarrow \chi \bar\chi $. In this section, we consider the simplified (and generic) case where the axion is thermalized in the early universe. This situation corresponds to $ T _{ a {\rm SM} } \lesssim  T _{ {\rm RH}} \lesssim   T _{ \chi  {\rm SM}} $ {\bf and} $ T _{ {\rm RH}} \gtrsim T _{ a \chi }$. This condition can only be satisfied for either the first or second hierarchy in Eq.~\eqref{eq:heirarchies}. 

As usual, the Boltzmann equation for the $ \chi $ number density is, 
\begin{equation}
\dot{n}_\chi + 3 H n_\chi = 2\bar{n}_{a} ^2 \left\langle \sigma _{ a a \rightarrow \chi \bar{\chi} } v \right\rangle  \,,
\end{equation}
where $\bar{n}_a$ is the number density of axions in thermal equilibrium. To calculate the collision term, we neglect the impact of Bose-enhancement and approximate the axion phase space density as a Maxwell-Boltzmann distribution with zero chemical potential and temperature $T^{\prime}$. Taking the thermal average of Eq.~\eqref{eq:aatochichi} we find the collision term, 
\begin{equation} 
\bar{n}_{a} ^2 \left\langle \sigma _{ a a \rightarrow \chi \bar{\chi} } v \right\rangle = \frac{1}{64\pi^5} T^{\prime} g_{a\chi}^4 \int_{4 m_\chi^2}^\infty ds\, K_1 \left ( \frac{\sqrt{s}}{T^{\prime}} \right ) \sqrt{s} [\tanh^{-1} (v_\chi) - v_\chi]\,.
\end{equation} 
Integrating over $T$, we find the yield
\begin{equation}
 Y_{\chi }\simeq \frac{1}{32\pi^5} g^4_{a\chi} \int_0^\infty dT^\prime \, \frac{1}{s_{\text{SM}}(T) H(T)} \int^\infty_{4 m_\chi^2} ds \, \sqrt{s} K_1 \left(\frac{\sqrt{s}}{T^\prime}\right) \left [\tanh^{-1} v _\chi  - v _\chi  \right] \,.
\label{eq:Yfreezein}
\end{equation}

We evaluate this equation numerically to get the freeze-in condition. Since this process is IR-dominated, we approximate $g_\star $ and $g_{\star,S} $ with their values at $ m _\chi $.
We find,
\begin{equation}
\Omega _\chi  \simeq \Omega _{ {\rm DM}}\left[ \frac{g_{\star,S} ( m _\chi ) }{g_{\star,S}(T_{a {\rm SM}})}\right]^{5/3}\left ( \frac{10 ^3 }{g_{\star,S}  ( m _\chi )\sqrt{g_\star ( m _\chi )}}  \right ) \left ( \frac{g_{a\chi}}{3\times 10^{-6}} \right )^4\,.
\label{eq:Omega-freezein}
\end{equation}
The factor in square brackets is present only when the production of dark matter occurs predominantly in the dark sector after decoupling from the SM, that is, for $m_\chi \lesssim T_{a {\rm SM}}$. We observe that the abundance is independent of $ m _\chi $. Furthermore, we conclude from Eq.~\eqref{eq:Omega-freezein} that the freeze-in process can generate the observed dark matter relic density for natural values of $ g _{ a \chi } \sim m _{ \chi } / f _a $ if $ m _\chi $ is near the weak scale. 

Having calculated the freeze-in abundance, we now calculate the constraints on freeze-in from measurements of $ \Delta N _{ {\rm eff}} $. The temperature of the dark sector can be tracked from an initial SM temperature, $ T _{a {\rm SM}} $,
\begin{equation} 
T^{\prime } = \left( \frac{g_{\star,S}(T)}{g_{\star,S}(T_{a {\rm SM}})} \right) ^{1/3} T\,, 
\label{eq:TprimeFI}
\end{equation} 
The absent factor of $ 1 +  ( 7 /8 ) g _\chi / g _a  $ in this equation relative to Eq.~\eqref{eq:TprimeFO} is because, in freeze-in, dark matter does not have a sizable energy density in the early universe. The corresponding $ \Delta N _{ {\rm eff}} $ is given by,
\begin{align}
    \Delta N_{\rm eff} & = \frac{1}{2} \frac{8  }{7}  \left( \frac{ 11 }{ 4 } \right) ^{ 4/3} g _a \left( \frac{ g _{ \star, S } ( T ) }{ g _{ \star, S } ( T _{a {\rm SM}} ) } \right) ^{ 4/3}\,, \\ 
& \simeq  \left( \frac{ 1.8 \times g _{ \star,S } ( T ) }{  g _{ \star ,S} ( T _{a {\rm SM}} )} \right) ^{ 4/3}\,. 
\end{align}
We plot this value in blue in Fig.~\ref{fig:Dark_Rad_Ti}. We conclude that current experiments restrict $ T _{a {\rm SM}} \gtrsim 150~{\rm MeV} $, while CMB-S4 will only be able to set a limit $T_{a {\rm SM}} \gtrsim 350~\text{MeV}$. From the expression for $ T _{ a {\rm SM}} $ as a function of $ f _a $ (Eq.~\eqref{eq:TaSM}), we observe that future measurements of $\Delta N_{\text{eff}}$ alone are unlikely to be able to probe freeze-in of DM from thermalized axions in the early universe.

\section{Conclusion}
\label{sec:conclusions}
The Strong CP problem is one of the strongest indicators for the existence of physics beyond the SM. The QCD axion is a simple extension to the SM and provides an elegant solution to the strong CP problem. This has lead to a widescale effort for its detection. If the axion has a decay constant $ f _a \lesssim 10 ^{ 11} ~{\rm GeV} $, then it is not expected to make up a significant component of dark matter. Nevertheless, the QCD axion may still play a critical role in setting the dark matter density if it acts as a mediator between dark matter and the SM. This is the question we set out to understand in this work. 

The dynamics of QCD axion-mediated dark matter depend crucially on whether either the axion, dark matter, or both, reach thermal equilibrium with the SM. We explore four different classes of thermal histories: dark matter freeze-out into SM particles, freeze-in from collisions of SM particles, dark sector freeze-out into axions, and dark sector freeze-in from axion collisions. These different possibilities span a large range of possible interaction strengths and reheating temperatures. We perform an exploratory study of the phenomenology, concluding that the thermal histories could be probed by a combination of astrophysical and cosmological probes. Furthermore, even though the axion interactions are dimension-5 operators, many viable thermal histories exist where the dynamics are largely insensitive to the reheating temperature. In the case of freeze-in, we find the QCD axion could have an interaction strength to dark matter of its natural size. These features make QCD axion-mediated dark matter a predictive and compelling framework which demands additional study. 

Our paper fills a gap in the literature by focusing on the intricacies associated with producing dark matter through a QCD axion portal. We showed that a vast collection of production mechanisms remain to be fully understood and probed in this context hence offering a rich landscape for axion phenomenology in the framework of dark matter detection.

\bigskip

\noindent
{\bf Note added --} During the preparation of this work we became aware of Ref.~\cite{OtherPaperDMAxion} which considers a similar scenario.

\acknowledgments
We thank Stefano Profumo for valuable insights about early universe axion physics and Aditya Parikh for clarifications on Sommerfeld enhancement in pseudoscalar mediated self-interacting dark matter. This research was supported in part by NSF CAREER grant PHY-1915852, in part by the U.S. Department of Energy grant number DE-SC0023093, and in part by the Office of High Energy Physics of the
U.S. Department of Energy under contract DE-AC02-05CH11231. Part of this work was performed at the Aspen Center for Physics, which is supported by National Science Foundation grant PHY-1607611. 

\appendix

\bibliographystyle{JHEP}
\bibliography{bibliography}

\end{document}